\documentclass[12pt]{article}

\usepackage{amsfonts}
\usepackage{bbm}

\newcommand{\bmat}{\left(\begin{array}}
\newcommand{\emat}{\end{array}\right)}
\newcommand{\uno}{\mathbbm{1}}
\def\NPB#1#2#3{Nucl. Phys. B{#1} (#2) #3}
\def\PLB#1#2#3{Phys. Lett. B{#1} (#2) #3}

\def\a{\alpha}
\def\b{\beta}
\def\g{\gamma}

\def\ep{\epsilon}

\def\-{\hphantom{-}}
\def\ov{\overline}
\def\s2{\frac{1}{\sqrt2}}

\def\oh{\frac{1}{2}}
\def\beq{\begin{equation}}
\def\eeq{\end{equation}}
\def\beqa{\begin{eqnarray}}
\def\eeqa{\end{eqnarray}}
\def\tr{{\rm tr \,}}
\def\Tr{{\rm Tr \,}}
\def\dim{{\rm dim \,}}
\def\T{{\rm T}}
\def\Z{{\mathbb Z}}
\def\diag{{\rm diag \,}}

\def\ca{{\cal A}}

\def\cg{{\cal G}}
\def\cn{{\cal N}}

\def\cz{{\cal Z}}
\def\car{{\cal R}}

\def\deq#1{\mbox{$D$=#1}}
\def\neq#1{\mbox{$\cn$=#1}}
\def\Dsl{\,\raise.15ex\hbox{/}\mkern-13.5mu D} 
\def\r#1{\mbox{{\bf #1}}}

\begin{document}
\pagestyle{empty}
\begin{flushright}
{\tt IFT-UAM/CSIC-04-24}
\end{flushright}
\vspace*{2cm}

\vspace{0.3cm}
\begin{center}
{\LARGE \bf Strings on Eight-Orbifolds }\\[1cm]
Anamar\'{\i}a Font
\footnote{On leave from Departamento de F\'{\i}sica, Facultad de Ciencias,
Universidad Central de Venezuela, A.P. 20513, Caracas 1020-A, Venezuela.}\\[0.2cm]
{\it  Instituto de F\'{\i}sica Te\'orica C-XVI,
Universidad Aut\'onoma de Madrid,\\[-1mm]
Cantoblanco, 28049 Madrid, Spain. }\\[0.3cm]
and \\[0.3cm]
Jos\'e Antonio L\'opez \\[0.2cm]
{\it  Centro de F\'{\i}sica Te\'orica y Computacional, Facultad de Ciencias, \\[-1mm]
Universidad Central de Venezuela,\\[-1mm]
A.P. 47270, Caracas 1041-A, Venezuela. }\\[1cm]
\normalsize{\bf Abstract} \\[3mm]
\end{center}

\begin{center}
\begin{minipage}[h]{15.5cm}
\normalsize{
We present several examples of $\T^8/P$ orbifolds with $P \subset SU_4$. 
We compute their Hodge numbers and consider turning on discrete torsion. 
We then study supersymmetric compactifications of type II, heterotic,
and type I strings on these orbifolds. Heterotic compactifications to
\deq2 have a $B$-field tadpole with coefficient given by that of the
anomaly polynomial. In the $SO_{32}$ heterotic with standard embedding the 
tadpole is absent provided the internal space has a precise value of the Euler 
number. Guided by their relation to type I, we find tadpole-free $SO_{32}$
heterotic orbifolds with non-standard embedding. 
}
\end{minipage}
\end{center}

\newpage

\setcounter{page}{1} \pagestyle{plain}
\renewcommand{\thefootnote}{\arabic{footnote}}
\setcounter{footnote}{0}

\section{Introduction}

F-theory compactified on certain Calabi-Yau four-folds leads to \neq1 four-dimensional 
vacua that can have phenomenological applications \cite{witnp}. Compactification
of M-theory and type IIA strings on the same manifolds gives related lower-dimensional
vacua that have some interesting features of their own \cite{vw, bb, svw, dm, gvw}.
In this note we study $\T^8/P$ orbifolds, with $P \subset SU_4$, that share many properties
of the smooth spaces and allow a simpler computation of the string spectra and
interactions \cite{dhvw}.

We consider Abelian point groups $P$ that can be products of up to three 
cyclic factors. We classify all crystallographic $\Z_N$ actions and also present
examples with more factors in which discrete torsion is allowed. The Hodge numbers 
$h_{p,q}$ of these orbifolds are efficiently computed using their connection to 
superconformal field theories. We exclude $P$'s that leave a sub-torus invariant,
so that $h_{0,1}=h_{0,3}=0$ but in general $h_{0,2} \not= 0$. 
In section \ref{basics} we review some basic properties of K\"ahler four-folds 
with vanishing first Chern class and collect some useful results that also
apply to our orbifolds.

Examples of compactification on $\T^8/P$ orbifolds have been described by several 
authors \cite{bch, sen96, gm, roy, acharya, drs}. In section 3 we explicitly 
compute the massless spectra of type II orbifolds  and show that they coincide sector 
by sector with known results obtained from Kaluza-Klein reduction \cite{dm, gukov, ggw, hlm}. 
We review this reduction in  order to include the case $h_{0,2} \not= 0$
that signals extended supersymmetry.

Heterotic strings generically have a $B$-field tadpole in \deq2 \cite{vw}. To
examine this problem we first consider compactification on smooth
Calabi-Yau four-folds with standard embedding. 
The full charged spectrum follows from  analyzing the zero modes of \deq10
gaugini whereas the number of net gauge singlet spinors can be determined from anomaly
factorization. We verify equality of the coefficient of the \deq2 anomaly polynomial
and the tadpole obtained by integration of the anomaly canceling term in \deq10,
as expected on general grounds \cite{genus}. The tadpole can only be canceled
in the $SO_{32}$ heterotic provided the four-fold has Euler number $\chi=180$.
Analogous results hold in orbifolds with standard embedding. When $h_{0,2} \not= 0$
tadpole cancellation requires $\chi=90(2+h_{0,2})$.

We are naturally led to look into heterotic orbifolds with general modular invariant
embeddings. We indeed find $SO_{32}$ models in which the full anomaly and thus
the tadpole vanishes. These examples are closely related to type I strings that are
our last topic. We discuss in some detail \deq2 type I vacua realized as type IIB 
orientifolds. Some models of this kind were constructed in \cite{fg}.
While these theories are necessarily free of gravitational and
gauge non-Abelian anomalies, they can have $U_1$ anomalies that are shown to be
canceled by exchange of RR scalars.

Section 4 is devoted to concluding remarks. The data of the $\T^8/P$ orbifolds
is collected in four tables found at the end of the paper.

\section{Eightfolds}
\label{basics}

\subsection{Calabi-Yau}

Calabi-Yau $d$-folds (${\rm CY}_d$'s) are of special interest because they admit a 
reduced number of covariantly constant spinors. For a ${\rm CY}_d$ the
holonomy group is strictly $SU_d$. In this section we collect some
useful facts about ${\rm CY}_4$'s that are characterized by Hodge numbers
$h_{p,q}$ satisfying $h_{p,q}=h_{q,p}$, $h_{p,q}=h_{4-p,4-q}$ and $h_{0,p}=h_{4-p,0}$.
Furthermore, $h_{0,0}=h_{0,4}=1$.
Recall that the Betti numbers are $b_n=\sum_{p+q=n} h_{p,q}$.

In general, a complex K\"ahler 4-fold $Y$ with vanishing first Chern class, $c_1[Y]=0$,
has holonomy inside $SU_4$. We assume that the manifold is neither
$\T^8$ nor products $\T^2 \times {\rm CY}_3$, $\T^4 \times {\rm K3}$
so that  $h_{0,1}= h_{0,3}=0$. The independent Hodge numbers are
$h_{0,2}$, $h_{1,1}$, $h_{1,2}$ and $h_{1,3}$. Using results given in \cite{klry}
one can show that 
\beq
h_{2,2}  =  2(22 + 10 h_{0,2} + 2 h_{1,1} + 2 h_{1,3} - h_{1,2}) \ .
\label{hdd}
\eeq
The Euler characteristic can thus be written as
\beq
\chi  =  6(8 + 4 h_{0,2} + h_{1,1} + h_{1,3} - h_{1,2}) \ .
\label{eulerc}
\eeq
Notice that $\chi$ is always multiple of six. When $h_{0,2}$ also
vanishes the manifold is a $\rm{CY}_4$. Another quantity of interest
is the signature $\tau=b_4^+ - b_4^-$, where $b_4^+$ and $b_4^-$ are the
number of self and antiself-dual harmonic 4-forms. Clearly,
$b_4^+ + b_4^-=2 + 2h_{1,3} + h_{2,2}$. {}From the expression of $\tau$ in terms
of Chern classes \cite{klry} we find
\beqa
b_4^- & = & 2 h_{1,3} + h_{1,1} - 2 h_{0,2} - 1  \nonumber \\
b_4^+ & = & 47 + 22 h_{0,2} + 4 h_{1,3} + 3 h_{1,1} - 2 h_{1,2} \ .
\label{b4mm}
\eeqa
In section \ref{2b} we will recover these results {} from supersymmetry and anomaly
cancellation in type IIB compactification on the four-fold. Notice that
$\tau= \frac{\chi}{3} + 32 + 16 h_{0,2}$.

Other useful results derived from general properties of four-folds 
with $c_1[Y]=0$ are
\beqa
\frac1{5760 (2\pi)^4} \int_Y  \tr R^4 + \frac54 (\tr R^2)^2 & = & 2 + h_{0,2} 
\nonumber \\[0.2cm]
\frac1{8 (2\pi)^4} \int_Y - \tr R^4 + \frac14 (\tr R^2)^2 & = & \chi \ ,
\label{pontry}
\eeqa
where $R$ is the curvature 2-form.

There is a family of $\rm{CY}_4$'s with structure 
${\rm K3} \times {\rm K3}/ \sigma$ where the involution $\sigma$
reverses the sign of the (2,0) forms of each K3 but leaves the
(4,0) form of the 4-fold invariant \cite{borcea}. The action of $\sigma$ on
each K3 is characterized by
two 3-ples $(r_i, a_i, \delta_i)$, $i=1,2$ and the Hodge numbers 
turn out to be
\beq
h_{1,1}  =  r_1 + r_2 + f_1 f_2 \quad ;  \quad
h_{1,2}  =  f_1 g_2 + g_1 f_2  \quad  ; \quad
h_{1,3}  =  40 - r_1 - r_2 + g_1 g_2 
\ ,
\label{vbo}
\eeq
where $f_i=1 + (r_i-a_i)/2$ and  $g_i=11 - (r_i+a_i)/2$. This is a 
generalization of the class of Voisin-Borcea $\rm{CY}_3$'s with
structure $\T^2 \times {\rm K3}/\hat \sigma$, where the involution
$\hat \sigma$ reverses the signs of the $\T^2$ (1,0) form
and the K3 (2,0) form. Since the K3's can be obtained by modding
$\T^4$ by a $\Z_M$ action we expect to find orbifolds whose Hodge
numbers coincide with (\ref{vbo}). Clearly we can also think of
4-folds that are elliptic fibrations of the form 
$\T^2 \times {\rm CY}_3/ \tilde \sigma$ where $\tilde \sigma$ reverses 
the signs of the $\T^2$ (1,0) and the ${\rm CY}_3$ (3,0) forms.

\subsection{Orbifolds}

For an orbifold $\T^{2d}/P$, the point group $P$ is the holonomy group. 
We consider Abelian point groups $P$ given by products of $\Z_N$ factors, with
each factor generated by a rotation $\theta$ that acts on the $\T^{2d}$ complex coordinates as 
\beq
\theta X_i = e^{2i\pi v_i} X_i \quad , \quad i=1, \cdots, d . 
\label{gact}
\eeq
Clearly $v_i={\rm int}/N$ for $\theta$ of order $N$. Existence of covariantly constant spinors,
that will  give unbroken supersymmetries upon string compactification on the orbifold, requires
\beq
\sum_i \, \pm v_i = 0 \, {\rm mod} \, 2 ,
\label{susycond}
\eeq
for some choice of signs. This is the condition $P \subset SU_4$. 
For definiteness we choose $\sum_i \,  v_i = 0$ and also demand that no 
sub-torus be left invariant. Then, there are at least two solutions of (\ref{susycond}).
If there are no extra solutions, the orbifold is a singular limit of a ${\rm CY}_d$ manifold.
Extra solutions could appear for $d \geq 4$, in particular for $d=4$ they actually imply $h_{0,2} > 0$. 

There is also a condition that $P$ acts crystallographically on the torus lattice $\Lambda$. 
The allowed Abelian actions can be found combining proper $\Z_N$ twists as explained
in \cite{ek}. In eight dimensions such proper twists exist for $N=15,16,20,24,30$
and have exponents $a_i/N$, where $a_i < N/2$ are integers relative prime to $N$.
In all cases, we can find supersymmetric twists. Except for $N=16$ they can be
realized on the $E_8$ root lattice by elements of the Weyl group, the $\Z_{30}$
being the Coxeter rotation. The $\Z_{16}$ rotation is the generalization of the
$\Z_4$ rotation acting on the $SO_4$ lattice, it can be simply realized on a 
hypercubic lattice with orthonormal basis.  

Combining the lower dimensional proper twists given in Table 1 of \cite{ek} it is a simple
exercise to find all other possible crystallographic inequivalent supersymmetric actions.
For example, for $\Z_N$ obtained in this way there is one solution for each $N=2,3,5,9,14,18$, 
and there are respectively 3,7,4,2,10,2, for $N=4,6,8,10,12,24$. 
Whenever allowed we take the  torus lattice $\Lambda$ to be the product of two-dimensional 
$SO_4$ or $SU_3$ root lattices. Some exceptions are the $\Z_5$, $\Z_8$, $\Z_9$ 
and $\Z_{14}$ for which $\Lambda$ can be taken to be respectively the product of two $SU_5$ 
root lattices, the hypercubic lattice, the $SU_9$ root lattice, and the $SO_{16}$ 
root lattice. In many cases one could also use the $E_8$ lattice. 
The inequivalent sets of $v_i$'s are either of type
$(\frac1{a}, -\frac1{a}, \frac1{b}, -\frac1{b})$, $a,b=2,3,4,6$;
$\frac1{N}(1,1,1,-3)$, $N=4,6$; $\frac1{N}(1,3,-2,-2)$, $N=6,8$, $\frac1{N}(1,3,-1,-3)$, $N=5,8,10$;
or are in the following list
\begin{center}
\footnotesize
\begin{tabular}{lllll}
$\frac16(1,1,2,-4)$ & $\frac1{12}(1,5,-1,-5)$ & $\frac1{12}(1,5,-3,-3)$ & $\frac1{15}(1,2,4,-7)$ & $\frac1{24}(1,5,7,-13)$ \\
$\frac18(1,5,-2,-4)$ & $\frac1{12}(1,5,1,-7)$ & $\frac1{12}(1,-5,-4,8)$ & $\frac1{16}(1,3,5,-9)$ & $\frac1{24}(3,9,-4,-8)$ \\
$\frac18(1,3,1,-5)$ & $\frac1{12}(1,-5,-2,6)$ & $\frac1{12}(2,4,-3,-3)$ & $\frac1{18}(1,7,-5,-3)$ & $\frac1{24}(3,9,-2,-10)$ \\
$\frac19(1,4,-2,-3)$ & $\frac1{12}(1,5,-2,-4)$ & $\frac1{14}(1,-3,-5,7)$ & $\frac1{20}(1,3,7,-11)$ & $\frac1{30}(1,-7,-11,17)$ 
\\
$\frac1{10}(1,-3,-2,4)$ & $\frac1{12}(1,-5,2,2)$. & & &  \\
\end{tabular}
\end{center}
Note that the irreducible twists are included.

Let us now briefly explain a simple procedure to compute the
cohomology of these toroidal orbifolds.  
To begin consider a (2,2) superconformal theory with $c=3d$ realized by $d$ 
free complex bosons $X_i$ and fermions $\psi_i$. It is well known that the 
ground states of the Ramond-Ramond (RR) sector are in one to one correspondence 
with the cohomology classes of $\T^{2d}$. Indeed, the Hodge numbers $h_{p,q}$ are given 
by the number of RR massless states with $(p,q)$ $U_1$ charges \cite{lvw}.

Next we quotient the theory by $P$ whose elements transform the fermions 
$\psi_i$ just as the bosons $X_i$, for instance as (\ref{gact}) for 
$\Z_N$ actions, so as to preserve the superconformal invariance.
As usual, the partition function will include a
sum over twisted sectors plus the orbifold projection. Schematically,
\beq
\cz = \frac1{|P|} \sum_{g \in P} \sum_{h \in P} \, \ep(g,h) \cz(g,h) .
\label{opf}
\eeq
The discrete torsion $\ep(g,h)$ satisfies $\ep(g,g)=1$
and furthermore \cite{vdt}
\beq
\ep(g, h_1h_2)  =  \ep(g,h_1) \ep(g,h_2) \quad ; \quad
\ep(g,h) \ep(h,g)  =  1 .
\label{epcon}
\eeq
For instance, when there are several $\Z_N$ factors there is the possibility 
of non-trivial discrete torsion. For  $\Z_N \times \Z_M$ generated by
$\a$ ($\a^N=1$) and $\b$ ($\b^M=1$)
\beq
\ep(\a^k \b^\ell, \a^s \b^t) = \ep_1^{kt -\ell s},
\label{disd}
\eeq
where $\ep_1^{gcd(N,M)}=1$. Similarly, for the product of three
factors with a third generator $\gamma$ ($\gamma^K=1$) we have
\beq
\ep(\a^k \b^\ell \g^m, \a^s \b^t \g^u) = \ep_1^{kt -\ell s}\,
\ep_2^{\ell u - mt} \, \ep_3^{k u - ms} ,
\label{dist}
\eeq
where $\ep_2^{gcd(M,K)}=1$ and $\ep_3^{gcd(N,K)}=1$.

From the partition function we can extract the spectrum of states.
In particular, for right-movers in the $\theta$-twisted sector the mass formula is
\beq
m^2(\theta) = \oh (r+v)^2 + N_R  + E_0 -\frac{c}{24} ,
\label{mfor}
\eeq
where $r$ is an $SO_{2d}$ weight that
arises from bosonization of the fermions and the twist vector $v$ 
has components $v_i$. $N_R$ is an oscillator number for
all bosons and $E_0$ is the shift in vacuum energy given by
\beq
E_0= \sum_i \oh |v_i|(1-|v_i|).
\label{vener}
\eeq
To each state we can associate the $U_1$ charge
\beq
q(r) = \sum_i r_i + \frac{c}6.
\label{uchar}
\eeq
Recall that $r$ is an spinorial weight in the R sector.
For left-movers there are analogous results. The number of
RR massless states with $(p,q)$ $U_1$ charges is the
Hodge number $h_{p,q}$ of $\T^{2d}/P$. Since we have chosen
$\sum_i v_i = 0$, it is easy to see that
there is only one state, appearing in the untwisted sector, with charges 
$(0,d)$. Hence, $h_{0,d}=1$ as it should for a space of holonomy inside $SU_d$.

To find the degeneracy of the RR ground states with $(p,q)$ charges
we simply use the orbifold projector obtained from the partition function 
\cite{fiq}. In Tables \ref{tab1}, \ref{tab2}, \ref{tab3} and \ref{tab4} we present 
the Hodge numbers for a few selected orbifolds. Some comments are in order.

\begin{trivlist}

\item[$\bullet$] 
All $\Z_N$ orbifolds with proper twist in eight dimensions, thus with
$N=15,16,20,24,30$, have the same Hodge numbers as the $\Z_{15}$ displayed
in Table \ref{tab1}.

\item[$\bullet$]
The first entry in Table \ref{tab2} corresponds to orbifold limits 
of ${\rm K3} \times {\rm K3}$.

\item[$\bullet$]
In Table \ref{tab2}, the $\Z_2 \times \Z_3$ and the $\Z_2 \times \Z_7$ are actually
$\Z_6$ and $\Z_{14}$. We choose to give the product realization because it shows the
structure $\T^2 \times {\rm CY}_3/ \tilde \sigma$. Similarly, some examples
in Table \ref{tab3} can be seen as $\Z_2 \times \Z_6$ or $\Z_3 \times \Z_6$.

\item[$\bullet$]
The first 3 examples in Table \ref{tab3} with $P= \Z_2 \times \Z_M \times \Z_2$ 
have the Borcea structure ${\rm K3} \times {\rm K3}/ \sigma$, with 
$\sigma$ given by the second $\Z_2$.
Indeed, the Hodge numbers coincide with (\ref{vbo}) upon taking
$r_1=18$, $a_1=4$ together with $r_2=18,14,16$, $a_2=4,6,8$ for $M=2,3,4$.
These models can also be regarded as elliptic fibrations
$\T^2 \times {\rm CY}_3/ \tilde \sigma$, with $\tilde \sigma$ given by the 
first $\Z_2$. Moreover, the ${\rm CY}_3$
is of Voisin-Borcea type $\T^2 \times K3/ \hat \sigma$ and the 
$r_2, a_2$ are precisely those that reproduce the ${\rm CY}_3$
Hodge numbers when the K3 is realized as $\T^4/\Z_M$ \cite{borcea, mv}.
The example with $P= \Z_2 \times \Z_2 \times \Z_3$ in Table
\ref{tab3} has Hodge numbers
that agree with (\ref{vbo}) for $r_1=18$, $a_1=4$, $r_2=10$ and $a_2=8$.

\item[$\bullet$]
It is easy to show that in all $\T^8/P$ Abelian orbifolds $h_{0,2}$ is
even as observed in the Tables. However, there is no such constraint
on $h_{0,2}$ in generic K\"ahler four-folds with vanishing first Chern class. 
For example, there is a smooth, simply-connected, manifold 
with $h_{0,2}=1$, $h_{1,1}=h_{1,3}=21$, $h_{1,2}=0$ and holonomy $Sp_2$. This manifold is
obtained by taking the quotient of ${\rm K3} \times {\rm K3}$ by the involution 
that exchanges both factors and then blowing-up \cite{sal}.

\item[$\bullet$]
In Table \ref{tab4} we collect some representative orbifolds with discrete torsion. 
The $\Z_2 \times \Z_2$ was described in detail in \cite{svw} as an example with 
negative Euler characteristic. As noticed in \cite{svw}, $\chi$ can become negative 
because discrete torsion tends to increase the value of $h_{1,2}$. 
The two $\Z_2 \times \Z_2 \times \Z_2$ were discussed in \cite{acharya} where it was shown 
that the mirror of the orbifold without discrete torsion is obtained taking $\epsilon_2=\epsilon_3=-1$. 
For a mirror pair $Y$, $Y^*$, $h_{p,q}(Y)=h_{4-p,q}(Y^*)$.

\item[$\bullet$]
All examples have $\chi$ multiple of 24. The Euler characteristic can be computed from 
the Hodge numbers or from the master orbifold formula of \cite{dhvw}. Likewise, $h_{2,2}$
can be found using (\ref{hdd}) or directly from the number of RR states with (2,2) charge.

\item[$\bullet$]
To properly discuss the resolution of singularities in these orbifolds is beyond the scope
of this paper. In the $\T^8/\Z_N$ with twist vector of the form $\frac1{N}(1,-1,1,-1)$,
$N=2,3,4,6$, or $\frac1{N}(1,3,-1,-3)$, $N=5,8,10$, the singularities are known to be of
terminal type \cite{morrist}. A property shared by orbifolds with terminal singularities
is that only the untwisted sector contributes to $h_{1,1}$. For the rest of the $\Z_N$
actions there is at least one twisted sector $\theta^k$ (and its inverse $\theta^{N-k}$) that
adds to $h_{1,1}$. The condition for the existence of such sectors,
when written in terms of the corresponding twist with components $kv_i$, 
coincides with the criterion that prevents the singularity from being terminal
\cite{reid, mmm}. Borrowing the notation of \cite{mmm}, the condition is
$\sum_i \langle kv_i \rangle \leq 1$, where $\langle kv_i \rangle \in [0,1)$ is
obtained from $kv_i$ by adding or subtracting one.

\end{trivlist}
 
\section{Compactification}
\label{comp}

To compute the massless fields of type II or heterotic strings 
compactified on a Calabi-Yau four-fold we can consider the ten-dimensional
bosonic fields whose zero modes are easy to count. Massless fermion
fields follow via supersymmetry but it is instructive, in particular
to reduce charged fields, to determine the fermionic zero modes.
To this end we use the same procedure applied in the case of
three-folds \cite{verde, phil}.

Let $\psi^c$ be an $SO_{1,9}$ Weyl spinor in the ${{\bf 16}_c}$.
We can write the $SO_{1,1} \times SO_8$ decomposition
\beq
\psi^c = \psi^+ \otimes \eta^s + \psi^- \otimes \eta^c \ ,
\label{psic}
\eeq
where $\psi^\pm$ are $\deq2$ Weyl spinors of $\pm$ chirality,
whereas $\eta^s$ and $\eta^c$ are $SO_8$ Weyl spinors. In
terms of spinorial weights $(\pm \oh, \pm \oh, \pm \oh, \pm \oh)$,
$\eta^s$ has even and $\eta^c$ odd number of $+$'s, notice that
each spinor is its own complex conjugate. When $\psi^c$ is
Majorana-Weyl, so are $\psi^\pm$. A spinor $\psi^s$ in the 
${{\bf 16}_s}$ of $SO_{1,9}$ has a decomposition analogous
to (\ref{psic}) with $\psi^+$ and $\psi^-$ exchanged.

In a K\"ahler four-fold of holonomy inside $SU_4$, spinors
can be written in terms of the covariantly constant spinors
and Dirac matrices that act as creation operators. Concretely,
\beq
\eta^s = \alpha \eta_+ 
+ \alpha_{\bar\imath \bar\jmath}\, \g^{\bar\imath} \g^{\bar\jmath}\, \eta_+
+ \alpha_{\bar\imath \bar\jmath \bar k \bar l} \,
\g^{\bar\imath} \g^{\bar\jmath} \g^{\bar k} \g^{\bar l} \, \eta_+ \ .
\label{etas}
\eeq
The indices $i, \bar\imath=1, \cdots, 4$, refer to the local
complex coordinates. The Dirac algebra is 
$\{\g^i, \g^j \}=\{\g^{\bar\imath}, \g^{\bar\jmath} \}=0$,
$\{\g^i, \g^{\bar\jmath} \}=2g^{i\bar\jmath}$, where $g^{i\bar\jmath}$
is the K\"ahler metric. The spinors $\eta_+$ and 
$\eta_- = \g^{\bar 1}\g^{\bar 2}\g^{\bar 3}\g^{\bar 4} \eta_+$
are covariantly constant, with $\eta_+$ characterized by
$\gamma^j \eta_+ =0$. The decomposition (\ref{etas}) reflects
the branching of the ${\textbf 8_s}$ of $SO_8$ into 
${\textbf 1} + {\textbf 6} + {\textbf 1}$ of $SU_4$. Similarly,
\beq
\eta^c = \alpha_{\bar\imath}\, \g^{\bar\imath} \eta_+
+ \alpha_{\bar\imath \bar\jmath \bar k} 
\g^{\bar\imath} \g^{\bar\jmath} \g^{\bar k}  \eta_+ \ ,
\label{etac}
\eeq
which corresponds to ${\textbf 8_c} = {\textbf 4} + \bar{\textbf 4}$.

The crucial fact in the expansions (\ref{etas}) and (\ref{etac})
is that the coefficients $\alpha_{\bar\imath \bar\jmath \cdots}$
are $(0,q)$ forms. Moreover, zero modes of the Dirac operator
are given by spinors whose coefficients are harmonic forms.
Thus, in a ${\rm CY}_4$ manifold, in which $h_{0,0}=h_{0,4}=1$
and other $h_{0,q}=0$, there are no zero modes from
$\eta^c$ and two zero modes from $\eta^s$, namely the $SU_4$ singlets
$\eta_+$ and $\eta_-$. {}From the
expansion (\ref{psic}) it follows that a massless dilatino
$\psi^c$ in $\deq10$ gives rise to two massless $\psi^+$'s in $\deq2$.
We will also allow for $h_{0,2} \not=0$ in which case the
number of $\psi^+$'s from $\psi^c$ is $\cn$, where
\beq
\cn=2+h_{0,2} \ .
\label{nsusy}
\eeq
The space-time components $\psi_\mu^c$ of a 
$\deq10$ Majorana-Weyl (M-W) gravitino $\psi_M^c$ also have an
expansion of type (\ref{psic}), with $\deq2$ spinors $\psi_\mu^{\pm}$.
Hence, the zero modes of $\eta^s$ give $\cn$ M-W gravitini 
$\psi_\mu^+$ of positive chirality in $\deq2$. Similarly, a $\deq10$ gravitino 
$\psi_M^s$ of opposite chirality gives $\cn$ gravitini $\psi_\mu^-$
of negative chirality in $\deq2$.

Expansions of the form (\ref{etas}) and (\ref{etac}) also apply 
to spinors with an extra holomorphic index. Both $\eta_i^s$ and
$\eta_i^c$ have coefficients that are $(1,q)$ forms and the
spinors are zero modes of the Dirac operator iff the forms are
harmonic. For the  complex conjugates $\eta_{\bar \imath}^s$ and
$\eta_{\bar \imath}^c$ the coefficients are $(q,1)$ forms. Thus, 
for example, the zero modes of $\psi_i^c$ and  $\psi_{\bar \imath}^c$ give
rise to $2h_{1,2}$ massless $\deq2$ M-W spinors of positive 
and $2(h_{1,1}+h_{1,3})$ of negative chirality. {} From $\psi_i^s$ and  
$\psi_{\bar \imath}^s$ we find an analogous result exchanging
chirality. 

Zero modes of bosonic fields are counted as usual. An $n$-form field gives $b_n$ 
massless scalars. Recall that in \deq2 modes with space-time indices have 
no degrees of freedom on-shell. 
The metric $G_{MN}$ gives the graviton and real scalars, $h_{1,1}$ from 
$G_{i\bar\jmath}$ and $2(h_{1,3}-h_{0,2})$ from $G_{ij}$ and $G_{\bar\imath\bar\jmath}$.
The need to subtract $2h_{0,2}$ is a general result \cite{Besse}, it 
can be simply seen in counting the number of scalars arising from the metric
in a $\T^{2d}$ compactification.

We use conventions such that \deq2 massless Weyl fermions of positive chirality are 
left-moving. Also, a left-moving massless scalar corresponds to a self-dual 1-form. 
With these results we turn below to determining the massless spectrum
of type II, heterotic and type I compactifications.
Type II compactification on smooth Calabi-Yau four-folds is well known \cite{dm, gukov, ggw, hlm}.
We will repeat the analysis in order to allow for the case $h_{0,2} \not = 0$ that is
common in orbifolds.

\subsection{Type IIA}
\label{2a}

In type IIA the two $\deq10$ gravitini have opposite chirality.
Our previous discussion shows that there is an equal number $\cn$
of positive and negative chirality gravitini in $\deq2$. Hence,
the resulting theory has $(\cn, \cn)$ supersymmetry. The gravitini
components with internal indices produce equal number 
$2(h_{1,1} + h_{1,3} + h_{1,2})$ of positive and negative chirality
Majorana-Weyl `modulini' in $\deq2$. The supersymmetric partners
arise from the bosonic fields. There are $(h_{1,1} + 2h_{1,3} -2h_{0,2})$
real scalars from the metric $G_{MN}$, $(h_{1,1} + 2h_{0,2})$ from the
NS-NS 2-form $B_{MN}$ and $2h_{1,2}$ from the R-R 3-form $C_{MNP}$.
There are no dynamical modes from the R-R 1-form. The dilaton
gives one scalar that belongs in the gravity multiplet together
with the $\cn$ dilatini of positive and negative chirality
and the non-dynamical metric and gravitini. 

The multiplets of $(\cn, \cn)$ supersymmetry in $\deq2$ can be
obtained by dimensional reduction of the multiplets of
$\deq4$ supersymmetry with $2\cn$ supercharges. The $\deq4$
chiral multiplet with $\cn$ real scalars and $\cn/2$ Majorana spinors
gives a $\deq2$ chiral multiplet with $\cn$ real scalars and equal
number $\cn$ of positive and negative chirality M-W
spinors. For $\cn=2,4,8$ the $\deq4$ vector multiplet has one vector, 
$(\cn-2)$ scalars and $\cn/2$ Majorana spinors, for $\cn=6$ the
content is the same as for $\cn=8$. Reducing the vector multiplet
gives a vector multiplet with the same content
in a chiral multiplet plus a non-dynamical vector. 
 
Given the fermionic and bosonic zero modes explained above, we
see that for $\cn=2,4,8$ there are  $2(h_{1,2} + h_{1,3})/\cn$
chiral multiplets and $2h_{11}/\cn$ vector multiplets
(with non-dynamical vectors arising from $C_{\mu i \bar\jmath}$).
For $\cn=6$ the multiplets are the same as for $\cn=8$.

Up to now we have neglected the effect of the $B$-field tadpole
that appears generically in IIA compactifications to \deq2 
\cite{vw,bb,svw}. This tadpole can be canceled by introducing a number
$n$ of fundamental strings given by
\beq
n= \frac{\chi}{24} - \frac{1}{8\pi^2} \int_{Y} dC \wedge dC \ .
\label{nfs}
\eeq
When $\chi/24$ is a positive integer the tadpole can be canceled,
without breaking supersymmetry and without fluxes, just by adding
this number of fundamental strings. Each string has a matter
content given by the light-cone world-sheet fields, i.e. eight
real scalars, eight positive and eight negative chirality M-W
fermions. Then, for example, for $\cn=2,4,8$, there will be
$\chi/3\cn$ further chiral multiplets. 

Type IIA compactification on orbifolds can be carried out explicitly. The case of
$\T^8/\Z_2$ was first presented in \cite{sen96}. For other orbifolds the massless states
can be found using the standard construction, explained for instance in the appendix
of \cite{fh}. As in \cite{sen96}, in many $g$-twisted sectors there are no RR massless 
scalars because the GSO projection forbids massless R states either for left-movers 
or right-movers. The exception is when $g$ leaves some direction unrotated, e.g. 
the untwisted sector or the $\beta$ sector in the $\Z_N \times \Z_M$ examples in Table 2.
In such sectors the RR states could still be eliminated by the orbifold projection.  

As it should, the number of massless chiral and vector multiplets in IIA orbifolds agrees 
with the general analysis for compactification on a smooth manifold, with the Hodge 
numbers computed in the orbifold sense as explained in section 2.2. The agreement is sector 
by sector. As an example, take the $\Z_3$ in Table 1. Together with $h_{2,2}^g$ and  
$\chi^g$ that will be needed in Type IIB and heterotic compactifications, the relevant Hodge 
numbers in sectors twisted by $g=\theta^n$ are
\beq
\begin{array}{cccccc}
g & h_{1,1}^g & h_{1,2}^g & h_{1,3}^g & h_{2,2}^g & \chi^g \\
\hline
\theta^0 & 8  & 4 & 8 & 18 & 54  \\
\theta + \theta^2 & 0  & 0 & 0 & 162 & 162  \\
\end{array}
\ .
\label{z3hodge}
\eeq
In type IIA on $\T^8/\Z_3$, with $h_{0,2}=4$ and $\cn=6$,  we find  3 chiral and 2 
vector multiplets in the untwisted sector, and no multiplets in the twisted sectors. 
As another example, take the $\Z_6$ in Table 1, with $h_{0,2}=0$, $\cn=2$ and
\beq
\begin{array}{cccccc}
g & h_{1,1}^g & h_{1,2}^g & h_{1,3}^g & h_{2,2}^g & \chi^g \\
\hline
\theta^0 & 8  & 2 & 0 & 18 & 30  \\
\theta + \theta^5 & 9  & 0 & 0 & 0 & 18  \\
\theta^2 + \theta^4 & 0  & 0  & 0 & 90 & 90  \\
\theta^3 & 6  & 10 & 5 & 24 & 6  \\
\end{array}
\ .
\label{z6hodge}
\eeq
Correspondingly, we find 2  chiral and 8
vector multiplets in the untwisted sector, 9 vector multiplets in $\theta + \theta^5$,
no multiplets in $\theta^2 + \theta^4$, and 15 chiral plus 6 vector in $\theta^3$.

\subsection{Type IIB}
\label{2b}

In type IIB the two $\deq10$ gravitini have the same chirality,
say ${{\bf 16}_c}$, so that they produce $2\cn$ positive chirality gravitini 
in $\deq2$. Hence, the resulting theory has $(2\cn,0)$ supersymmetry. 
The gravitini components with internal indices produce M-W spinors, 
$N(\psi^-)=4(h_{1,1} + h_{1,3})$ of negative and 
$N(\psi^+)=4h_{1,2}$ of positive chirality.
The dilaton and the axion give one scalar each, from the metric
and the two antisymmetric tensors there are $(3h_{1,1} + 2h_{1,3} + 2h_{0,2})$
scalars. {} From the R-R 4-form $C_{MNPQ}$ with self-dual field
strength, there arise $h_{1,2}$ non-dynamical vectors from components
$C_{\mu i \bar\jmath \bar k}$ and $C_{\mu \bar \imath j k}$, and massless scalars
{} from components with four internal indices. The self-dual harmonic 4-forms
give scalars that are left-moving or positive chirality. Hence,
there are  $b_4^+$ scalars of positive and $b_4^-$ of negative chirality.

Excluding the dilaton that belongs to the gravity multiplet, the total
number of positive and negative chirality scalars are
\beqa
N(\varphi^+) & = & 3h_{1,1} + 2 h_{1,3} + 2 h_{0,2} + 1 + b_4^+ \nonumber \\[0.2ex]
N(\varphi^-) & = & 3h_{1,1} + 2 h_{1,3} + 2 h_{0,2} + 1 + b_4^- \ .
\label{npsis}
\eeqa
By supersymmetry $N(\varphi^-) = N(\psi^-)$ and therefore $b_4^-$ must be given by
(\ref{b4mm}). Also, using the formula for $b_4^+$ readily gives $N(\varphi^+)=\chi + 4 h_{1,2}$.

We now discuss anomaly cancellation as in \cite{dm}.
Using the conventions of \cite{verde}, the contributions of
$\psi_\mu^+$, $\psi^+$ and $\varphi^+$ to the anomaly polynomial in
$\deq2$ are respectively
\beq
I_{3/2}^+= -\frac{23}{48} \tr R^2 \quad , \quad
I_{1/2}^+= \frac{1}{48} \tr R^2 \quad , \quad
I_{0}^+= \frac{1}{24} \tr R^2  \ .
\label{acontri}
\eeq
The gravity multiplet that includes $2\cn(\psi_\mu^+, \psi^-)$ contributes
$I_{grav}= -{\cn} \tr R^2$.
Then, the total anomaly is
\beq
\frac{1}{24} \tr R^2 \left[ -24 \cn + 2 h_{1,2} -2(h_{1,3} + h_{1,1}) 
+ b_4^+ - b_4^- \right] \ ,
\label{tota}
\eeq
which cancels by virtue of (\ref{b4mm}).

Compactification of type IIB on $\T^8/P$ orbifolds can also be carried out in detail \cite{fh}.
However, since the theory is chiral, care has to be taken because in the light-cone space-time
is not directly visible. For NS states this is not a problem since, for instance, basically 
only $\psi_{-\oh}^m |0 \rangle$ ($r=(\underline{\pm 1,0,0,0})$ in the notation of \cite{fh})
gives rise to massless dynamical degrees of freedom in NSNS or NSR sectors. For both right and
left-moving R states we must
instead relax the GSO projection in a way that a decomposition such as (\ref{psic}) for 
\deq10 spinors be manifest. This means that R states in the light-cone can be {\it both}
of type $\eta^s$ (i.e. $r$ with $\sum_a r_a={\rm even}$ in the notation of \cite{fh})
or type $\eta^c$ ($\sum_a r_a={\rm odd}$). For a ${{\bf 16}_c}$ M-W the former corresponds
to positive and the latter to negative chirality \deq2 M-W spinors. This modification
is also necessary for the R sector in heterotic orbifolds and in open strings in orientifolds.
In type IIB when we combine right and left-moving R states, the tensor product $\psi^+ \otimes \psi^-$
does not give physical fields (it would be part of a vector) but $\psi^+ \otimes \psi^+$
and $\psi^- \otimes \psi^-$ give respectively positive and negative chirality scalars.   

As it should, the type IIB orbifold spectra agree with the general analysis for 
smooth manifolds. In each sector twisted by $g$ we find
$N^g(\varphi^-) = N^g(\psi^-)= 4(h_{1,1}^g+ h_{1,3}^g)$, $N^g(\psi^+)=4 h_{1,2}^g$ and
$N^g(\varphi^+)=\chi^g + 4 h_{1,2}^g$. 

\subsection{Heterotic}
\label{hetsec}

In heterotic compactification to \deq2 there is a potential $B$-field tadpole
arising from the \deq10 action term that cancels the anomaly, namely
$\int B \wedge X_8$ \cite{vw}. In the notation of \cite{verde}, the 8-form $X_8$ is
\beq
8X_8 = \frac1{3} \Tr_A F^4 - \frac1{900} (\Tr_A F^2)^2 -\frac1{30} \Tr_A F^2 \tr R^2 + \tr R^4
+ \frac14 (\tr R^2)^2 \ .
\label{eform}
\eeq
With a convenient normalization the coefficient of the tadpole is 
\beq
c= \frac1{48 (2\pi)^4} \int_Y X_8 \ .
\label{tadhet}
\eeq
Clearly, $c$ could vanish when the gauge fields have a precise background. 

An alternative way of determining $c$ is to calculate the anomaly polynomial
of the \deq2 theory obtained upon compactification. This anomaly must be
exactly of the form 
\beq
\ca=c\left( \tr R^2 - v_a \tr F_a^2 \right ) \ ,
\label{fullarf}
\eeq
where, using conventions specified below, $v_a=1$ for all group factors.
The point is that this anomaly would be canceled through the Green-Schwarz mechanism
precisely by a term $c\int B$ in the \deq2 action. In the case of the standard embedding
we will compute $c$ in both ways and show agreement. To this end we 
first determine the massless fields upon reduction.

In the heterotic string there is one $\deq10$ gravitino,
chosen in the ${{\bf 16}_c}$, that produces 
$\cn$ positive chirality gravitini in $\deq2$.  
Hence, the resulting theory has $(\cn,0)$ supersymmetry. 
The dilatino in the ${{\bf 16}_s}$ gives $\cn$ negative chirality 
M-W spinors that also belong in the gravity multiplet.
The gravitino components with internal indices produce M-W spinors, 
$2(h_{1,1} + h_{1,3})$ of negative and 
$2h_{1,2}$ of positive chirality. From the metric
and the antisymmetric tensor there are $2(h_{1,1} + h_{1,3})$
scalars. 

We next perform the reduction of the $\deq10$ gauge multiplet
with group $\cg$ equal to $E_8 \times E_8$ or $SO_{32}$.
We will assume that $h_{0,2}=0$ so that the holonomy is exactly
$SU_4$. We can then realize the standard embedding, i.e. choose a
background for the gauge connection equal to the spin connection.
This background breaks $\cg$ to the commutant $G$ that gives a maximal
subgroup $G \times SU_4 \subset \cg$. Then, the standard embedding
gives group $SO_{10} \times E_8$ or $SO_{24} \times U_1$. 

To determine the massless fermions in various representations we
look for zero modes of the $\deq10$ 
gaugini $\lambda_\a$ that are also M-W in the ${{\bf 16}_c}$. 
The standard embedding implies that $SU_4$ gauge transformations 
are identified with $SU_4$ holonomy transformations. This means
for instance that the internal Dirac operator acting on 
gaugini that are $SU_4$ singlets is just like the internal Dirac
operator acting on neutral spinors. We already know that
there are two zero modes $\eta_+$ and $\eta_-$ in this case.
Therefore, $SU_4$ singlet gaugini just give rise to two positive chirality
gaugini of the unbroken group $G$ in $\deq2$. 

To study $SU_4$ charged gaugini we need to decompose 
the $\cg$ adjoint under $G\times SU_4$. For $SO_{10} \times SU_4 \subset E_8$
we have  
\beq
{\bf 248} = ({\bf 45}, {\bf 1}) \oplus ({\bf 16}, {\bf 4}) \oplus
({\bf \ov {16}}, {\bf \ov{4}}) \oplus ({\bf 10}, {\bf 6}) 
\oplus ({\bf 1}, {\bf 15}) \ .
\label{ae8}
\eeq
We see that there are gaugini, transforming in the $SO_{10}$ adjoint,
that are $SU_4$ singlets and that were already discussed in 
the previous paragraph.
There are also components, denoted $\lambda_i$, that transform in 
the {\bf 16} of $SO_{10}$ and the {\bf 4} of $SU_4$.
Due to the $SU_4$ index $i$, zero modes of $\lambda_i$
are just like zero modes of the gravitino $\psi_i$, and similarly
for the $\lambda_{\bar \imath}$ in the ${\bf \bar{4}}$. 
Hence, there are $(h_{1,1} + h_{1,3})$ negative chirality and $h_{1,2}$
positive chirality spinors transforming in ${\bf 16} + {\bf \ov{16}}$
of the unbroken $SO_{10}$. Since the {\bf 6} of $SU_4$ is the antisymmetric
product of two fundamentals, we can argue that for the components transforming in 
the {\bf 10} of $SO_{10}$ and the  {\bf 6} of $SU_4$  the internal spinors in the
decomposition of type (\ref{psic}) have two antisymmetric holomorphic indices
so that the zero modes correspond to $(2,q)$ harmonic forms. This then gives 
$2h_{1,2}$ negative chirality and $h_{2,2}$
positive chirality spinors transforming in ${\bf 10}$. We will
see that this result is consistent with anomaly factorization.
We can also check it explicitly in orbifold examples. For the
components transforming as {\bf 15} of $SU_4$ we will just assume
that they give a net  number $N_{0G}$ of positive chirality
spinors singlets of $SO_{10}$.

To compute the anomaly polynomial we need the contribution of fermions
transforming in a representation $\car$ of the unbroken gauge group.
For positive chirality this is
\beq
I_{1/2}^+= \frac{\dim \car}{48} \tr R^2 - \frac{1}{2} \Tr_{\car} F^2  \ .
\label{fcontri}
\eeq
In general we can write $\Tr_{\car} F^2 = T(\car) \tr F^2$.
We use conventions such that for $E_8$, $T({\bf 496})= 30$,
for $SU_N$, $T(\car)=\frac12, \frac{N-2}2, N$, for the fundamental, 2-index
antisymmetric and adjoint,
for $SO_{2N}$, $T(\car)=1, 2^{N-4}, (2N+2), (2N-2)$, for the vector, spinor,
2-index symmetric and adjoint. 
The full anomaly polynomial takes the form (\ref{fullarf}), 
only the overall constant $c$ is model dependent. 
In compactifications of the $E_8 \times E_8$ heterotic on a ${\rm CY}_4$
with standard embedding, necessarily $c=30$ since the unbroken group
includes the hidden $E_8$. The same coefficient for the
observable $SO_{10}$ follows, using (\ref{hdd}) for $h_{2,2}$, 
provided that the numbers of spinors transforming as ${\bf 10}$
are exactly as claimed before. Requiring that the gravitational anomaly 
also appears with $c=30$ fixes the net number of singlet spinors 
arising from the \deq10 gaugini. We find 
\beq
N_{0G} = 510 - \chi \ .
\label{nsing}
\eeq
Recall that there are spinors arising from the gravitino that are obviously 
gauge singlets. Taking these into account, the total net number of
positive chirality singlet spinors is $N_{0T} = 526 - 4\chi/3$.
This result can be verified in explicit orbifold examples.
  
To determine the spinors arising from gaugini in the $SO_{32}$ heterotic 
compactified on a ${\rm CY}_4$ with standard embedding, we need the decomposition of
the adjoint ${\bf 496}$ under $SO_{24}\times U_1 \times SU_4$. This is
\beq
{\bf 496} = ({\bf 276},0, {\bf 1}) \oplus ({\bf 1}, 0, {\bf 1}) \oplus
({\bf 24}, q, {\bf 4}) \oplus
({\bf 24}, -q,{\bf \ov{4}}) \oplus ({\bf 1}, 2q, {\bf 6}) \oplus
({\bf 1}, -2q, {\bf 6}) \oplus ({\bf 1},0, {\bf 15}) \ ,
\label{aso}
\eeq
where $q=\frac{1}{2\sqrt2}$ in order that the $U_1$ generator $Q$
has the correct normalization $\Tr Q^2 = 30$.
According to our previous discussion the $SU_4$ background
breaks $SO_{32}$ to $SO_{24} \times U_1$ with the following
charged spinor content:  2 positive chirality gaugini transforming in the adjoint,
$(h_{1,1} + h_{1,3})$ negative chirality and $h_{1,2}$
positive chirality spinors transforming in $({\bf 24},q) + ({\bf 24},-q)$,
as well as $2h_{1,2}$ negative chirality and $h_{2,2}$
positive chirality spinors transforming in $({\bf 1},2q) + ({\bf 1},-2q)$. Moreover, there must be a net number 
$N_{0G}$ of positive chirality
spinor singlets $({\bf 1},0)$. 
The $SO_{24}$ and $U_1$ gauge pieces of the full anomaly polynomial are easy to
compute. Using (\ref{eulerc}) and (\ref{hdd}) both give $c=30-\frac{\chi}6$.
The gravitational piece appears with this same $c$ provided that $N_{0G}$
is given by (\ref{nsing}), as expected since this value is a property
of the four-fold.

It remains to analyze the zero modes of the \deq10 gauge vectors. 
Besides the (non-dynamical) vectors partners of the \deq2 gaugini, by supersymmetry
we expect to obtain scalars to pair into chiral multiplets with the negative
chirality spinors. For instance, in the $E_8 \times E_8$ heterotic there must be
$(h_{1,1} + h_{1,3})$ real scalars transforming in ${\bf 16} + {\bf \ov{16}}$  and $2h_{1,2}$
in ${\bf 10}$. It is useful to organize the fields into multiplets of $(2,0)$ supersymmetry
that can be obtained decomposing those of (2,2).
The gauge multiplet contains gauge vectors and two positive chirality M-W spinors.
The chiral multiplet, denoted generically $\Phi_2^-$, contains one complex scalar and one
negative chirality Weyl spinor. The so called Fermi multiplet \cite{witn2} contains
one positive chirality Weyl spinor, denoted generically $\Psi^+$. Then, for example, 
the (2,0) theory arising upon compactification of the $E_8 \times E_8$ heterotic
on a ${\rm CY}_4$ has the following multiplets
\beq
\Phi_2^- [ (h_{1,1} + h_{1,3})(\r{16} + \r1) + h_{1,2} \r{10} ]
+ \Psi^+ [ \frac{h_{2,2}}2 \r{10} + \frac{ N_{0G}}2 \r1 ] \ ,
\label{fullsod}
\eeq
where we have not included the $S0_{10}\times E_8$ gauge multiplets. All
fields in (\ref{fullsod}) are $E_8$ singlets. For the $SO_{32}$ heterotic
the observable group is $S0_{24}\times U_1$ and in (\ref{fullsod})
we just need to replace $\r{16}$ by $(\r{24},q)$, $\r{10}$ by $(\r1,2q)$
and $\r1$ by $(\r1,0)$. These results are reproduced in orbifold compactifications
discussed shortly.

Now that we have computed the tadpole coefficient from the anomaly we wish
to show that the same result follows from (\ref{tadhet}). The task is basically to 
determine the adjoint traces $\Tr_A F^2$ and $\Tr_A F^4$ in the standard embedding.
To this end we use the adjoint decompositions, (\ref{ae8}) or (\ref{aso}),
our $SU_4$ conventions for $\Tr_{\car} F^2$ and the relations (see the appendix
of \cite{ksty})
$$
\Tr_{\bf 4} F^4 = \frac3{16} (\tr F^2)^2 - \frac14 \tr F^4 \quad ; \quad
\Tr_{\bf 6} F^4 = \tr F^4 \quad ; \quad
\Tr_{\bf 15} F^4 = 3(\tr F^2)^2 - 2\tr F^4 \ .
$$
Furthermore, since $\tr R^2$ is evaluated in the ${\textbf 8_v} = {\textbf 4} + \bar{\textbf 4}$,
we find that in the standard embedding $\tr F^2 = \tr R^2$ while 
$\tr F^4 = \frac34 (\tr R^2)^2 - 2 \tr R^4$. Collecting all intermediate results we arrive at
\beq
X_8^{E_8 \times E_8} = \frac18 \tr R^4 + \frac5{32} (\tr R^2)^2 \quad ; \quad
X_8^{SO_{32}} = \frac98 \tr R^4 - \frac3{32} (\tr R^2)^2 \ .
\label{xochos}
\eeq
Using (\ref{pontry}) to perform the integration we readily find
$c(E_8 \times E_8)=30$ and $c(SO_{32})=30-\frac{\chi}6$, in accordance
with the anomaly computation. Thus, with standard embedding only the
perturbative $SO_{32}$ heterotic could be tadpole-free provided the
four-fold has $\chi=180$. Such four-folds do exist \cite{cy4}. For
instance, there are 28 of them among the transversal hypersurfaces in 
weighted projective space. The example of lowest degree and reflexive
weights is a hypersurface in ${\mathbb P}_5^{(9,9,11,18,22,30)}[99]$ having
$h_{1,1}=27$, $h_{1,2}=60$ and $h_{1,3}=55$. 

It is also interesting to discuss four-folds with $h_{0,2}\not=0$. For example, for
${\rm K3} \times {\rm K3}$ with $h_{0,2}=2$, the holonomy group is $SU_2 \times SU_2$.
The standard embedding then gives group $SO_{12} \times E_8$ or $SO_{24} \times U_1^2$.
In the $E_8 \times E_8$ heterotic matter can be neutral or transform in the spinorial
${\bf 32}$ or ${\bf 32'}$ or vector ${\bf 12}$ of the commutant $SO_{12}$.
Based on the branching of the $E_8$ adjoint representation
under $SO_{12} \times SU_2 \times SU_2$ we find
$\oh(h_{1,1} + h_{1,3})$ negative chirality and $\oh h_{1,2}$
positive chirality M-W spinors transforming in ${\bf 32} + {\bf 32'}$, as well as
$2h_{1,2}$ negative chirality and $(h_{2,2}-4)$
positive chirality spinors transforming in ${\bf 12}$. {}From
anomaly factorization we find the total net number of positive chirality
singlet spinors (including modulini) to be $N_{0T}=1016-8\chi/3$.  
These same results are obtained in orbifold realizations of ${\rm K3} \times {\rm K3}$
as well as in other $h_{0,2}=2$, $\Z_N$ or $\Z_N \times \Z_M$ orbifolds, with or
without discrete torsion. The $SO_{32}$ heterotic is just as simple to work out.
In general, in orbifolds with $h_{0,2} \not= 0$ and standard embedding
the tadpole coefficients turn out to be $c(E_8 \times E_8)=15(2+h_{0,2})$ and 
$c(SO_{32})=15(2+h_{0,2})-\frac{\chi}6$. Instead oh having to choose an internal
space with a precise $\chi$ an obvious alternative to cancel the tadpole is to make a different
embedding and this can be most easily carried out in orbifold compactifications.

Compactification of heterotic strings on $\T^8/P$ orbifolds, in particular computation of 
the massless spectrum, follows as usual \cite{sierra, fh} and taking into account
the slight modification explained at the end of section \ref{2b}.
To begin we need to specify the embedding
in the gauge degrees of freedom. We use the bosonic formulation
and realize a $\Z_N$ rotation (\ref{gact}) by a shift vector $V$ such that
$NV \in \Gamma$, where $\Gamma$ is either the $E_8 \times E_8$
or the $Spin(32)/\Z_2$ lattice. Modular invariance of the partition
function requires
\beq
N(V^2 - v^2) = 0\ {\rm mod}\  2 \ .
\label{famod}
\eeq
The standard embedding $V=(v_1,v_2,v_3, v_4,0,\cdots, 0)$ trivially satisfies (\ref{famod})
but there are many other solutions.

The orbifolds with $h_{0,2}=0$ are presumably singular limits of 
${\rm CY}_4$'s. The simplest example to compare with smooth compactification is 
the $\Z_4$  in Table 1 with
\beq
\begin{array}{cccccc}
g & h_{1,1}^g & h_{1,2}^g & h_{1,3}^g & h_{2,2}^g & \chi^g \\
\hline
\theta^0 & 16  & 0 & 0 & 36 & 72  \\
\theta + \theta^3 & 16  & 0 & 0 & 0 & 32  \\
\theta^2 & 0  & 0 & 0 & 136 & 136  \\
\end{array}
\ .
\label{z4hodge}
\eeq
The shift $V$ for the standard embedding breaks $SO_{32}$ to $SO_{24} \times U_1 \times SU_4 $.
We find the following (2,0) massless matter multiplets:
\beq
\begin{array}{ll}
\theta^0 : & 3 \Psi^+ (\r1, 2q, \r6) + 
 4\Phi_2^-[(\r{24},q,\r4) + 4(\r1,0,\r1)] \\[0.5ex]
\theta+ \theta^3: & 16\Phi_2^-[(\r{24},q,\r1) + 4(\r1,0,\r4)]\\[0.5ex]
\theta^2: &  \Psi^+ [68(\r1,2q,\r1) + 60(\r1,0,\r6)]  \\
\end{array}
\ ,
\label{spz4ht}
\eeq
where $q=1/2\sqrt2$.
Using the data in (\ref{z4hodge}) we can check that in the twisted sectors the number 
of $SO_{24} \times U_1$ charged multiplets  agrees with the analysis for smooth 
${\rm CY}_4$'s, c.f. (\ref{fullsod}).
Agreement in the untwisted sector requires counting $SU_4$ dimensionality 
as multiplicity, i.e. assuming that $SU_4$ is completely broken. 
Among the massless states there are candidate Higgs fields, namely
the scalars that are inert under $SO_{24} \times U_1$ and sit at each fixed point in the 
$\theta+ \theta^3$ sectors. These orbifold states are indeed of blowing-up type 
since they have left-moving oscillators acting on the twisted vacuum \cite{hv}. 
We also find that the total net number of positive chirality spinors is
$N_{0T} = 526 - 4\chi/3$ provided that we include the $SU_4$ gaugini.
It is straightforward to do the standard embedding for other ${\rm CY}_4$ orbifolds
of the $E_8 \times E_8$ or the $SO_{32}$ heterotic string. 
Some examples were studied in \cite{bch}.
In all cases there are
scalars that can break the observable group to $SO_{10}$ or $SO_{24} \times U_1$.
 
We now wish to present an example with vanishing tadpole in 
the $SO_{32}$ heterotic. For the $\Z_3$ of Table 1 we take the non-standard embedding 
\beq
V=\frac13(1, \cdots, 1,0,0,0,0,0,0) \ ,
\label{vz3}
\eeq
which leaves $SU_{10} \times SO_{12} \times U_1$ unbroken. The massless spectrum is
\beq
\begin{array}{ll}
\theta^0 : & (\Phi_6^- + \Psi^+)[(\r{10},\r{12}, \frac1{2\sqrt5}) + (\r{45},\r1, -\frac1{\sqrt5}) +4(\r1,\r1,0)] 
\\[0.5ex]
\theta + \theta^2: &  81\Psi^+ (\r1,\r1,\frac{\sqrt5}{3}) \ ,
\end{array}
\label{z3ori}
\eeq
where $\Phi_6^-$ stands for a
(6,0) chiral multiplet that includes four complex scalars and four negative chirality Weyl spinors.
It is easy to check the vanishing of the full anomaly polynomial.
Another distinctive feature is the absence of non-Abelian charged matter
in the twisted sectors, an important fact in identifying this orbifold as dual to an orientifold 
studied in the next section.

Taking hints from the orientifolds of next section we have found other 
$SO_{32}$ non-standard embeddings leading to models with $c=0$. There is one such model
for each $\Z_N$ with $N$ odd. The corresponding shifts and unbroken group are
\beq
\begin{array}{ccc}
N & V & G \\
\hline
5 &  ({\frac15}^6, {\frac25}^6, 0^4) & U_6^2 \times SO_8 \\
9 & ({\frac19}^4,{\frac29}^4,{\frac49}^4,{\frac39}^2, 0^2) & U_4^3 \times U_2 \times SO_4 \\
15 & ({\frac1{15}}^2,{\frac2{15}}^2,{\frac4{15}}^2,{\frac8{15}}^2,{\frac3{15}}^2{\frac6{15}}^2,{\frac5{15}}^2, 0^2)
& U_2^7 \times SO_4 
\end{array}
\ ,
\label{vcnq}
\eeq
where the notation, e.g. ${\frac15}^6$ means $\frac15$ repeated six times.
The massless spectra are straightforward to compute but too cumbersome to display.
In all cases we find that the full anomaly polynomial, including $U_1$ terms, does vanish.

\subsection{Type I}
\label{unosec}

It is natural to study type I compactifications that will give $(\cn,0)$ theories
with a charged sector arising from D$p$-branes. These vacua can be described as \deq2
type IIB orientifolds. 
These orientifolds must be free of gravitational and non-Abelian anomalies. There could
only be $U_1$ anomalies that can be canceled by couplings of RR scalars to the
$U_1$ field strength. We will show explicit examples having these properties.

We follow the construction 
used in \cite{gp, d6} for \deq6 and in \cite{afiv} for \deq4. The orientifold
group is $G_1 + \Omega G_2$, we focus mostly on $G_1=G_2=\Z_N$. The closed string
states are those of type IIB on $\T^8/\Z_N$ invariant under $\Omega$. The
open string states depend on the matrices $\gamma_{g,p}$ that
realize the $\Z_N$ action on the Chan-Paton factors. These matrices are constrained
by the orientifold group structure and tadpole cancellation.

Tadpoles are particularly easy to compute for $N$ odd. In this case cancellation
requires D9-branes only, $\gamma^T_{\Omega,9}=\gamma_{\Omega,9}$ and 
\beq
\Tr \gamma_{2k,9} = 32 \prod_{j=1}^4 \cos k\pi v_j  \quad ; \quad k=0,1, \cdots,N-1 \ .
\label{tadodd}
\eeq   
It is also consistent to take $\gamma_{1,9}^N=\uno$. As a simple example consider the
$\Z_3$ of Table 1. The solution to (\ref{tadodd}) is
\beq
\gamma_{1,9} = \diag(e^{2i\pi/3} \uno_{10}, e^{-2i\pi/3} \uno_{10}, \uno_{12}) \ .
\label{z3gamma}
\eeq
The resulting gauge group is $SU_{10} \times SO_{12} \times U_1$.
The open string spectrum includes exactly the same charged fields appearing in the
untwisted sector of the heterotic orbifold (\ref{z3ori}). This is not surprising
since in the auxiliary shift formalism \cite{afiv}, (\ref{z3gamma}) is equivalent to (\ref{vz3}).
The closed string spectrum is
\beq
\begin{array}{ll}
\theta^0 : & 4(\Phi_6^- + \Psi^+) (\r1,\r1,0)
\\[0.5ex]
\theta + \theta^2: &  81\phi^+ (\r1,\r1,0)  \\
\end{array}
\ .
\label{z3oricl}
\eeq
In the untwisted sector there appear precisely the neutral 
multiplets needed to complete the heterotic untwisted sector. In the twisted
sectors instead of the positive chirality spinors found in the heterotic,
there appear positive chirality scalars. Upon fermionization
the number of fields is the same and the total gravitational, as well
as the gauge non-Abelian anomalies are still identically zero. 
The orientifold $U_1$ is anomalous because the $\phi^+$'s are not charged.
However, based on general arguments the anomaly can be compensated by anomalous
$U_1$ transformations of the $\phi^+$ that are RR scalars \cite{iru}.

In \deq2 the type I and the heterotic dilaton are identical \cite{abpss} so that
examples of exact weak/weak duals in which the perturbative spectra
coincide are expected. This can occur in \deq4 as well \cite{abpss, kks}.
In the other \deq2 odd $\Z_N$ orientifolds the 99 sector fully matches 
the charged untwisted multiplets of a perturbative $SO_{32}$ heterotic with embedding given by 
the auxiliary shift. The untwisted closed states complete the heterotic untwisted sector. 
In $\Z_5$ and $\Z_{15}$ the heterotic twisted states are all non-Abelian 
singlets and match the twisted closed states up to fermionization and $U_1$ charges.
In $\Z_9$ the heterotic $\theta^3, \theta^6$ sectors include some non-Abelian
charged fields that could be Higgsed away as in \cite{kks}.
The gauge group is of the form $\prod_\a U_n^\a \times SO_{2\ell}$,
where $\a=1, \cdots, \frac{(N-1)}2$. The auxiliary shift $V$ has components $V^\a$ repeated $n_\a$
times plus $\ell$ zeroes,  see (\ref{vz3}) and (\ref{vcnq}).
The $U_1^\a$ in each $U_n^\a$ is anomalous but the anomaly can be compensated by
transformation of RR scalars. 

We can adapt the analysis of \cite{iru} to verify cancellation of the 1-loop anomaly
that in \deq2 comes from the vacuum polarization diagram. In the closed string channel
the relevant counter diagram is the annulus with RR fields propagating along and one $U_1^\a$
coupled at each boundary. The contribution to the anomaly coming from this graph is
proportional to
\beq
A_\a = \frac{n_\a}{N} \sum_{k=1}^{N-1} C_k(v) \sin^2 2\pi k V^\a  \ ,
\label{ancoef}
\eeq
where the coefficients $C_k(v)$ are
\beq
C_k(v) = \prod_{i=1}^{4} 2 \sin k\pi v_i   \ .
\label{ckcoef}
\eeq
In all cases we find that $A_\a$ equals the coefficient of the 1-loop anomaly,
namely $-\oh \Tr Q_\a^2$ with normalization such that $Q_\a$ charges are multiples
of $1/\sqrt{2n_\a}$. Mixed $U_1^\a U_1^\b$ anomalies cancel in  similar fashion.
Cancellation of $U_1$ anomalies in models with D1-branes at ${\mathbb C}^4/\Z_N$ 
singularities through the same mechanism was discussed in \cite{mohri}.

In addition to D9-branes, even order orientifolds require D5-branes and/or
D1-branes. A simple example with only D5-branes is the $\Z_4$ generated by
$\frac14(2,-2,1,-1)$. Tadpole cancellation implies 99 group $U_8^2$. With
all D5-branes at the origin the 55 group is the same. The matter content
is somehow similar to that in the \deq6, $\Z_4$ orientifold \cite{d6}. 
Examples with D5 and D1-branes were analyzed in \cite{fg}.

We now wish to discuss an orientifold with D1-branes and to this end consider the
$\Z_2$ of Table 1 whose generator will be denoted $R$. 
This example was first studied in \cite{fg}.
The structure of tadpoles
is very similar to that in the GP orientifold \cite{gp}. There are the usual tadpoles
proportional to $V_2 V_8$ that require 32 D9-branes ($V_2$ and $V_8$ are respectively
the regularized space-time volume and the torus volume). There are also
tadpoles proportional to $V_2/V_8$ that require 32 1-branes. Furthermore,
\beqa
\gamma^T_{\Omega,9}=\gamma_{\Omega,9} & ; & \gamma^T_{\Omega R,1}=\gamma_{\Omega R,1} \\
\gamma^T_{\Omega,1}=\gamma_{\Omega,1} & ; & \gamma^T_{\Omega R,9}=\gamma_{\Omega R,9} \ ,
\label{gatra}
\eeqa
where the first line is needed for tadpole cancellation and the second follows
because the GP action is such that $\Omega^2=\uno$ also on 19-states. Finally, there
are the twisted tadpoles proportional to $V_2$ and to
\beq
\sum_{I=1}^{256} \left( \Tr \gamma_{R,9} + 16 \Tr \gamma_{R,1,I} \right)^2 \ .
\label{twitad}
\eeq
Cancellation of (\ref{twitad}) implies $\Tr \gamma_{R,9}=\Tr \gamma_{R,1,I}=0$.
$I$ runs over the fixed points of $R$.

Without loss of generality we can take $\gamma_{\Omega,9}=\gamma_{\Omega,1}=\uno$.
The algebra then implies that $\gamma_{R,9}=\gamma_{\Omega R,9}$ and 
$\gamma_{R,1}=\gamma_{\Omega R,1}$ so that the matrices $\gamma_{R,p}$ are
symmetric and traceless. We can choose
\beq
\gamma_{R,9} = \diag(\uno_{16}, -\uno_{16}) \quad ; \quad 
\gamma_{R,1,I} = \diag(\uno_{m_I}, -\uno_{m_I}) \,
\label{z2gam}
\eeq
where $2m_I$ is the number of 1-branes at fixed point $I$.

The gauge group for 99 strings is $SO_{16} \times SO_{16}$ and the matter
content is an (8,0) chiral multiplet $\Phi_8^-$ transforming in $(\r{16}, \r{16})$ 
($\Phi_8^-$ contains 4 complex scalars and 4 negative chirality Weyl spinors).
For $1_I 1_I$ strings the gauge group is $SO_{m_I} \times SO_{m_I}$ and matter is 
a $\Phi_8^-$  in (\textbf {\emph{m}${}_I$}, \textbf {\emph{m}${}_I$}). 
There can also be $m_J$ D1-branes sitting at a non-fixed point $J$ (and thus the 
same number at the $\Z_2$ image). Then, $\sum_I m_I + \sum_J m_J=16$.
The gauge group for $1_J 1_J$ strings is $SO_{m_J}$ since the $R$ projection only 
exchanges $J$ and its image, and the $\Omega$ projection just implies anti-symmetric 
Chan-Paton factor. There is also a $\Phi_8^-$ transforming in 
$\frac{{\textbf {\emph{m}${}_J$}}({\textbf {\emph{m}${}_J$}} +\bf 1)}{\bf 2}$, 
the $\Omega$ projection implies symmetric Chan-Paton because of the extra sign 
due to DD boundary conditions.   

For 19 strings inspection of the cylinder partition function reveals no massless
states in the NS sector and only one positive chirality M-W spinor in the R sector.
This spinor is invariant under the $\Z_2$ rotation so the $R$ projection implies
invariant Chan-Paton factor. The representations are then
(\r{16}, \r1;\textbf {\emph{m}${}_I$}, \r1) plus (\r1, \r{16};\r1,\textbf {\emph{m}${}_I$})
for $9 1_I$ strings and for $9 1_J$, 
(\r{16}, \r1;\textbf {\emph{m}${}_J$}) plus (\r1, \r{16};\textbf {\emph{m}${}_J$}).

Finally, the closed strings provide the (8,0) gravity multiplet and the moduli,
i.e. eight singlet chiral multiplets, {} from the untwisted sector. 
The $\Omega$ projection removes all states
{}from the twisted sector. Given the full gauge and matter content
it is easy to check that the gravitational and gauge anomalies do vanish.
 
Unlike the GP model, now there is no way to cancel tadpoles locally so we do not
expect to find a weak heterotic dual. A very simple  configuration, with group
just $SO_{16} \times SO_{16}$ {}from 99 strings, has 16 wandering D1-branes.
Matter is comprised by one chiral multiplet in (\r{16}, \r{16}) from 99 strings,
sixteen singlet chiral multiplets from 11 strings,
sixteen positive chirality M-W spinors in (\r{16}, \r1) plus (\r1,\r{16}) from $91_J$,
and the moduli from the closed untwisted strings.

It is rather easy to find a $V$ giving $SO_{16} \times SO_{16}$ upon
heterotic compactification on $\T^8/\Z_2$. In both heterotics $V$ is such that 
there is no massless matter in the twisted sector. In the untwisted sector we find the 
moduli together with one charged chiral multiplet, transforming as (\r{16}, \r{16}) and as
\mbox{(\r{128}, \r1) + (\r1, \r{128})} in the $SO_{32}$ and the $E_8 \times E_8$
heterotic respectively. The two heterotics have total anomaly polynomial of the
form (\ref{fullarf}) with $c=-8$. It is tempting to suppose that
in the $SO_{32}$ the tadpole can be canceled by adding sixteen 1-branes, dual to the 
wandering D1-branes in the orientifold. In the $E_8 \times E_8$
the tadpole could be canceled by adding wrapped M2-branes.

When $c < 0$, adding strings dual to wandering D1-branes can offset the tadpole 
because each brane by itself has an anomaly polynomial of the form (\ref{fullarf}) 
with coefficient $c_1=\frac12$. Indeed, recall that from 11 strings the degrees of 
freedom are eight real scalars and eight M-W spinors of negative chirality whereas 
from 19 strings there is one positive chirality M-W spinor originally transforming 
in the vector of $SO_{32}$.

\section{Conclusions}

Consistent compactifications to \deq2 do belong in the full landscape of string vacua
and deserve further study. Constructing $\T^8/P$ orbifolds and using them to compactify
strings is a natural project that, to our knowledge, had not been previously
carried out systematically. In this paper we have partially classified such orbifolds 
with $P \subset SU_4$. We have also explained how to adapt the usual rules to compute the 
orbifold spectrum of states to the \deq2 case that requires care for chiral theories.

We have noticed that the $B$-field tadpole generically present in heterotic compactifications
is most easily computed as the coefficient of the anomaly polynomial. In the perturbative
heterotic string this coefficient follows directly from the massless spectrum. This gives a 
simple prescription to search for perturbative tadpole-free models. We have discovered a few examples 
in the $SO_{32}$ heterotic. Those with standard embedding require an internal space with
$\chi=90(2+ h_{0,2})$. We have not found this Euler number in orbifolds but rather in
Calabi-Yau four-folds. The tadpole-free models with non-standard embedding are connected
to type IIB $\Z_{odd}$ orientifolds. In fact, analyzing orientifolds in which the anomaly
is always absent suggests possible ways to cancel the heterotic tadpole via non-perturbative
effects, for instance turning on strings dual to wandering D1-branes.

Among possible applications of our results we can envisage the study of the effective \deq2 
theories. For example, as in \cite{gukov, hlm, gh} one could consider type IIA compactifications
on  $\T^8/P$ orbifolds. Finally, let us remark that M-Theory and F-theory compactifications on these
orbifolds can be explored as well.


\vspace*{1cm}

\noindent
{\bf Acknowledgments : } A.F. is grateful to: S. Theisen for truly useful conversations
and instructive comments; M. Kreuzer for indicating the 4-folds with $\chi=180$; and the 
Albert Einstein Institute as well as the Erwin Schr\"odinger Institute for hospitality
and support while working on this paper. J.A.L. thanks Fonacit for a doctoral fellowship.

\vspace*{2cm}

\begin{table}[htb]
\footnotesize
\renewcommand{\arraystretch}{1.25}
\begin{center}
\begin{tabular}{|c|c|c|c|c|c|c|c|}
\hline
$P$  &  Generator &  $h_{0,2}$ & $h_{1,1}$ & $h_{1,2}$ & $h_{1,3}$ &
$h_{2,2}$ & $\chi/24$   \\    \hline\hline
$\Z_2$ & $(\oh, -\oh, \oh, -\oh)$ & 6 &  16 & 0 & 16 & 292 & 16 \\
[0.2ex] 
\hline
$\Z_3$ & $(\frac13, -\frac13, \frac13, -\frac13)$ & 4 & 8 & 4 & 8 & 180 &
9 \\ [0.2ex]
\hline 
$\Z_N$, $N=4,6$ & $(\frac1{N}, -\frac1{N}, \frac1{N}, -\frac1{N})$ & 4 & 8 & 0 & 8 & 188 &
10 \\ [0.2ex]
\hline
$\Z_N$, $N=4,6$ & $(\frac12, -\frac12, \frac1{N}, -\frac1{N})$ & 2 & 16 & 16 & 16 & 180 &
8 \\ [0.2ex]
\hline 
$\Z_5$ & $(\frac15, \frac35, -\frac15, -\frac35)$ & 2 & 4 & 4 & 4 & 108 &
5 \\ [0.2ex]
\hline
$\Z_8$ & $(\frac18, \frac38, -\frac18, -\frac38)$ & 2 & 4 & 0 & 4 & 116 &
6 \\ [0.2ex]
\hline
$\Z_4$ & $(\frac14, \frac14, \frac14, -\frac34)$ & 0 & 32 & 0 & 0 & 172 &
10 \\ [0.2ex]
\hline
$\Z_6$ & $(\frac16, \frac16, \frac26, -\frac46)$ & 0 & 23 & 12 & 5 & 132 &
6 \\ [0.2ex]
\hline
$\Z_9$ & $(\frac19, \frac49, -\frac29, -\frac39)$ & 0 & 15 & 11 & 0 & 82 &
3 \\ [0.2ex]
\hline
$\Z_{15}$ & $(\frac{1}{15}, \frac{2}{15}, \frac{4}{15}, -\frac{7}{15})$ & 0 & 8 & 0 & 0 & 76 &
4 \\ [0.2ex]
\hline
\end{tabular}
\end{center}
\caption{ $\T^8/\Z_N$ \label{tab1} }
\end{table}

\begin{table}[htb]
\footnotesize
\renewcommand{\arraystretch}{1.25}
\begin{center}
\begin{tabular}{|c|c|c|c|c|c|c|c|}
\hline
$P$  &  Generators &  $h_{0,2}$ & $h_{1,1}$ & $h_{1,2}$ & $h_{1,3}$ &
$h_{2,2}$ & $\chi/24$   \\    \hline\hline
$\Z_N \times \Z_M$ & $(\frac1{N}, -\frac1{N}, 0, 0)$ & 2 & 40 & 0 & 40 & 404 &
24 \\
$N, M=2,3,4,6$ & $(0,0,\frac1{M}, -\frac1{M})$ & & & & & &  \\ [0.2ex]
\hline
$\Z_2 \times \Z_3$ & $(\oh, -\oh, -\oh,\oh)$ & 2 & 16 & 16 & 16 & 180 & 8  \\
{} & $(0,0,\frac13, -\frac13)$ & & & & & & \\ [0.2ex] 
\hline
$\Z_2 \times \Z_3$ & $(\oh, -\oh, 0,0)$ & 0 & 32 & 21 & 5 & 150 & 6  \\
{} & $(0,\frac13,\frac13, -\frac23)$ & & & & & & \\ [0.2ex] 
\hline
$\Z_2 \times \Z_3$ & $(\oh, -\oh, -\oh, \oh)$ & 0 & 28 & 13 & 1 & 134 & 6  \\
{} & $(0,\frac13,\frac13, -\frac23)$ & & & & & & \\ [0.2ex] 
\hline
$\Z_2 \times \Z_4$ & $(\oh, -\oh, 0,0)$ & 0 & 54 & 16 & 2 & 236 & 12 \\
{} & $(0,\frac14,\frac14, -\frac12)$ & & & & & & \\ [0.2ex]
\hline
$\Z_2 \times \Z_4$ & $(\oh, -\oh, \oh, -\oh)$ & 0 & 64 & 0 & 8 & 332 & 20  \\
{} & $(0,\frac14,\frac14, -\oh)$ & & & & & & \\ [0.2ex] 
\hline
$\Z_2 \times \Z_7$ & $(\oh, -\oh, -\oh, \oh)$ & 0 & 16 & 9 & 1 & 94 & 4  \\
{} & $(0,\frac17,\frac27, -\frac37)$ & & & & & & \\ [0.2ex] 
 \hline
\end{tabular}
\end{center}
\caption{ $\T^8/\Z_N \times \Z_M$\label{tab2} }
\end{table}

\begin{table}[htb]
\footnotesize
\renewcommand{\arraystretch}{1.25}
\begin{center}
\begin{tabular}{|c|c|c|c|c|c|c|c|}
\hline
$P$  &  Generators &  $h_{0,2}$ & $h_{1,1}$ & $h_{1,2}$ & $h_{1,3}$ &
$h_{2,2}$ & $\chi/24$   \\    \hline\hline
$\Z_2 \times \Z_2 \times \Z_2$ & $(\oh, -\oh, 0,0)$ & 0 & 100 & 0 & 4 & 460 &
28  \\
{} & $(0,0,\frac12, -\frac12)$ & & & & & & \\
{} & $(0,\frac12, -\frac12,0)$ & & & & & & \\ [0.2ex] \hline
$\Z_2 \times \Z_3 \times \Z_2$ & $(\oh, -\oh, 0,0)$ & 0 & 72 & 8 & 8 & 348 &
20 \\
{} & $(0,0,\frac13, -\frac13)$ & & & & & & \\
{} & $(0,\frac12, -\frac12,0)$ & & & & & & \\ [0.2ex] \hline
$\Z_2 \times \Z_4 \times \Z_2$ & $(\oh, -\oh, 0,0)$ & 0 & 118 & 0 & 2 & 524 &
32 \\
{} & $(0,0,\frac14, -\frac14)$ & & & & & & \\
{} & $(0,\frac12, -\frac12,0)$ & & & & & & \\ [0.2ex] \hline
$\Z_2 \times \Z_2 \times \Z_3$ & $(\oh, -\oh, 0,0)$ & 0 & 44 & 16 & 12 & 236 &
12 \\
{} & $(0,0,\frac12, -\frac12)$ & & & & & & \\
{} & $(0,\frac13, -\frac13,0)$ & & & & & & \\ [0.2ex] \hline
$\Z_2 \times \Z_3 \times \Z_3$ & $(\oh, -\oh, 0,0)$ & 0 & 96 & 21 & 5 & 406 &
22 \\
{} & $(0,0,\frac13, -\frac13)$ & & & & & & \\
{} & $(0,\frac13, -\frac13,0)$ & & & & & & \\ [0.2ex] \hline
$\Z_3 \times \Z_3 \times \Z_3$ & $(\frac13, -\frac13, 0,0)$ & 0 & 220 & 0 & 0 &
924 & 57  \\
{} & $(0,0,\frac13, -\frac13)$ & & & & & & \\
{} & $(0,\frac13, -\frac13,0)$ & & & & & & \\ [0.2ex]
\hline
\end{tabular}
\end{center}
\caption{ $\T^8/\Z_N \times \Z_M \times \Z_K$  \label{tab3} }
\end{table}

\begin{table}[htb]
\footnotesize
\renewcommand{\arraystretch}{1.35}
\begin{center}
\begin{tabular}{|c|c|c|c|c|c|c|c|c|}
\hline
$P$  &  Generators & Phases &  $h_{0,2}$ & $h_{1,1}$ & $h_{1,2}$ & $h_{1,3}$ &
$h_{2,2}$ & $\chi/24$  \\    \hline\hline
$\Z_2 \times \Z_2 $ & $(\oh, -\oh, 0,0)$ & $\ep_1=-1$ & 2 & 8 & 64 &
8 & 20 & -8\\
{} & $(0,0,\frac12,  -\frac12)$ &  & & & & & & \\ [0.2ex] \hline
$\Z_3 \times \Z_3 $ & $(\frac13, -\frac13, 0,0)$ & $\ep_1=e^{2i\pi/3}$ & 2 & 4 &
36 & 4 & 44  & -3   \\
{} & $(0,0,\frac13, -\frac13)$ & & & & & & & \\ [0.2ex] \hline
$\Z_4 \times \Z_4 $ & $(\frac14, -\frac14, 0,0)$ & $\ep_1=-1$ & 2 & 24 & 0 & 24 & 276 & 16   \\
{} & $(0,0,\frac14, -\frac14)$ & $\ep_1=i$ & 2 & 4 & 16 & 4 & 84 & 2 \\ [0.2ex] \hline
$\Z_2 \times \Z_2 \times \Z_2$ & $(\oh, -\oh, 0,0)$ & $\ep_1=1$, $\ep_3=-1$ & 
& & & & & \\
{} & $(0,0,\frac12, -\frac12)$ & $\ep_2=1$ & 0 & 20 & 64 & 20 & 76 & -4\\
{} & $(0,\frac12, -\frac12,0)$ & $\ep_2=-1$ & 0 & 4 & 0 & 100 & 460 & 28 \\ [0.2ex] \hline
$\Z_2 \times \Z_2 \times \Z_3$ & $(\oh, -\oh, 0,0)$ & $\ep_1=-1$ & 0 & 20 & 48 & 4 & 44 & -4 \\
{} & $(0,0,\frac12, -\frac12)$ &  & & & & & & \\
{} & $(0,\frac13, -\frac13,0)$ &  & & & & & &  \\ [0.2ex] \hline
$\Z_3 \times \Z_3 \times \Z_3$ & $(\frac13, -\frac13, 0,0)$ & $\ep_1=1$, $\ep_3=e^{2i\pi/3}$ &
& & & & & \\
{} & $(0,0,\frac13, -\frac13)$ & $\ep_2=1$ & 0 & 22 & 63 & 9 & 42 & -6\\
{} & $(0,\frac13, -\frac13,0)$ & $\ep_2=e^{2i\pi/3}$ & 0 & 4 & 0 & 36 & 204 & 12 \\ [0.2ex] 
\hline
\end{tabular}
\end{center}
\caption{ Examples with discrete torsion \label{tab4} }
\end{table}

\end{document}